\documentclass[aip, amsmath,amssymb, reprint]{revtex4-1}
\usepackage{graphicx}
\usepackage{dcolumn}
\usepackage{bm}
\usepackage{tabularx}

\usepackage[utf8]{inputenc}
\usepackage[T1]{fontenc}
\usepackage{mathptmx}
\usepackage{color, soul}
\usepackage{siunitx}
\usepackage{soul,xcolor}
\usepackage{lineno}
\usepackage{comment}
\graphicspath{ {Figures/} }

\newcommand{\ba}{\begin{eqnarray}}
\newcommand{\ea}{\end{eqnarray}}

\begin{document}
\setstcolor{red}
\preprint{AIP/123-QED}

\title{Performance limits due to thermal transport in graphene single-photon bolometers}

\author{Caleb Fried}%
\altaffiliation{These authors contributed equally to this work}
\affiliation{Department of Electrical Engineering and Computer Science, Massachusetts Institute of Technology, Cambridge, MA 02139}
\affiliation{Department of Earth and Planetary Sciences, Harvard University, Cambridge, Massachusetts 02138, USA}
\affiliation{Department of Physics, Harvard University, Cambridge, Massachusetts 02138, USA}
\affiliation{Raytheon BBN, Quantum Engineering and Computing Group, Cambridge, Massachusetts 02138, USA}
\author{B. Jordan Russell}%
\altaffiliation{These authors contributed equally to this work}
\affiliation{Department of Physics, Washington University in St.\ Louis, St.\ Louis, MO, USA}
\author{Ethan G. Arnault}%
\affiliation{Department of Electrical Engineering and Computer Science, Massachusetts Institute of Technology, Cambridge, MA 02139}
\author{Bevin Huang}%
\affiliation{Intelligence Community Postdoctoral Research Fellowship Program, Massachusetts Institute of Technology, Cambridge, MA 02139}
\author{Gil-Ho Lee}%
\affiliation{Department of Physics, Pohang University of Science and Technology, Pohang 790-784, Republic of Korea}
\author{Dirk Englund}%
\affiliation{Department of Electrical Engineering and Computer Science, Massachusetts Institute of Technology, Cambridge, MA 02139}
\author{Erik A. Henriksen}%
\affiliation{Department of Physics, Washington University in St.\ Louis, St.\ Louis, MO, USA}
\author{Kin Chung Fong}%
 \email{kc.fong@rtx.com}
\affiliation{Department of Physics, Harvard University, Cambridge, Massachusetts 02138, USA}
 \affiliation{Raytheon BBN, Quantum Engineering and Computing Group, Cambridge, Massachusetts 02138, USA}

\date{\today}

\begin{abstract}
In high-sensitivity bolometers and calorimeters, the photon absorption often occurs at a finite distance from the temperature sensor to accommodate antennas or avoid the degradation of superconducting circuitry exposed to radiation. As a result, thermal propagation from the input to the temperature readout can critically affect detector performance. In this report we model the performance of a graphene bolometer, accounting for electronic thermal diffusion and dissipation via electron-phonon coupling at low temperatures in three regimes: clean, supercollision, and resonant scattering. Our results affirm the feasibility of a superconducting readout without Cooper-pair breaking by mid- and near-infrared photons, and provide a recipe for designing graphene absorbers for calorimetric single-photon detectors. We investigate the tradeoff between the input-readout distance and detector efficiency, and predict an intrinsic timing jitter of $\sim$2.7 ps. Based on our result, we propose a spatial-mode-resolving photon detector to increase communication bandwidth.
\end{abstract}

\maketitle

\section{Introduction}
Bolometers are high-sensitivity direct detectors of electromagnetic radiation by sensing the temperature rise of a material resulting from photon absorption. Their remarkable sensitivity and broad spectral bandwidth have found numerous applications across various fields, including radio-astronomy, spectroscopy, and the observation of rare events like dark matter interactions\cite{Pirro.2017,Schütte-Engel.2021}. More recently, bolometers have demonstrated an alternative approach to the traditional heterodyne technique for measuring superconducting qubits\cite{Gunyhó.2023}. When bolometers achieve single-photon sensitivity and function as calorimeters, they can provide distinct advantages by playing a crucial role in creating entangled quantum states over long distances\cite{Couteau.2023}. The ability to establish entanglement in a broad electromagnetic spectrum using single-photon bolometers can open up exciting possibilities for advancing quantum communication and information processing technologies over many different hardware platforms.

Graphene is a nearly ideal material for bolometers\cite{Ryzhii.2013,Cai.2014,Fatimy.2016,Sassi.2017,Zoghi.2019,Blaikie.2019,Xie.2019,Miao.2018,Skoblin.2018,Haller.2022,Chen.2022} with potential for single-photon sensitivity\cite{Vora.20128z,Yan.2012ldkp,Fong.2012,Lee.2020,Kokkoniemi.2020,Seifert.2020,Battista.2022} and use in investigating mesoscopic thermodynamics\cite{Walsh.2017,Lee.2020,Pekola.2022}. The electronic heat capacity of graphene can be exceptionally small due to its minute density of states, which can approach one Boltzmann constant, $k_B$, at low temperatures\cite{Vora.20128z,Fong.2012,Aamir.2021} that allow for a substantial temperature rise from absorbing one photon. As a zero-band gap semi-metal, graphene can absorb photons across an extensive range of the electromagnetic spectrum, spanning from ultraviolet to microwave frequencies\cite{Sarma.2011}. Meanwhile, the electron-phonon (e-ph) coupling in graphene is very weak\cite{Hwang.2008,Bistritzer.2009,Kubakaddi.2009,Viljas.2010,Song.2012,Chen.2012,Massicotte.2021} due to the requirement of momentum conservation in a small Fermi surface. Finally, the electron-electron (e-e) interaction time is so fast---approaching the Heisenberg uncertainty limit---that heat is rapidly and efficiently diffused among the electrons\cite{Lucas.201738q,Gallagher.2019,Block.2021}. This characteristic is pivotal for both maintaining well-defined localized temperatures within the material and preserving a significant portion of the photon energy in the electrons by mitigating losses to optical phonons during the hot electron cascade\cite{Song.201102t,Tielrooij.2013}. Everything considered, we can attribute these desirable bolometric properties to the unique linear band structure of graphene, which distinguishes it from other materials commonly used in bolometers.

\begin{figure}\centering
\includegraphics[width=\columnwidth]{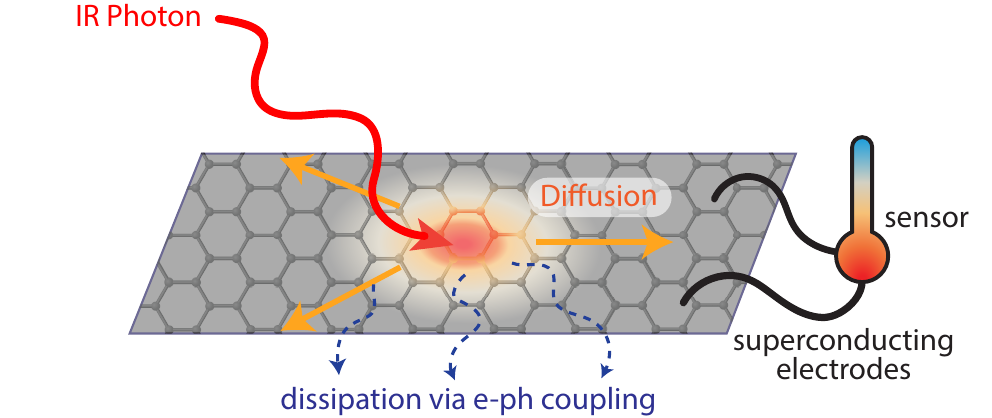}
\caption{\textbf{Schematic of an infrared graphene bolometer.} The bolometer comprises a graphene sheet and a thermal sensor readout. When a single infrared photon is absorbed in the graphene, the electron temperature is elevated in a localized hotspot. Heat spreads by thermal diffusion and ultimately dissipates into the lattice via e-ph coupling. A superconductor-based sensor can measure the temperature rise, but must be isolated from the photon input to avoid degradation by photon-induced Cooper pair breaking.}
\label{fig:concept}
\end{figure}

Nevertheless, achieving graphene single-photon bolometers remains a substantial challenge\cite{Du.2014} because of the short thermal relaxation time of graphene electrons\cite{Efetov.2018} and the lack of fast and direct readout of the electron temperature\cite{Han.2013,Fatimy.2016,Zoghi.2019}, $\Theta_e$. Graphene-superconductor hybrid components have been developed to address these obstacles, including Josephson junctions\cite{Walsh.2017,Lee.2020,Walsh.2021}, superconducting resonators\cite{Kokkoniemi.2020,Katti.2022}, and normal metal-insulator-superconductor tunnel junctions\cite{Vora.20128z}. But their integration to the photon input is not straightforward because the energy carried by infrared (IR) photons exceeds the superconducting gap, leading to the breaking of Cooper pairs which degrades the bolometer performance. One particular crucial design consideration is the distance separating the superconducting readout elements and the location of incoming photon (see Fig.\ 1). As with transition edge sensors, the thermal properties of the photon absorber can strongly impact the performance of a single-photon bolometer, e.g.\ jitter, detection speed, dead time, efficiency, and dark count. Previously, the heat diffusion and dissipation in graphene was studied with continuous heating or high optical fluence in the pump-probe experiments at higher temperatures \cite{Ruzicka.2010,Song.201102t,Brida.2013,Jago.2019,Draelos.2019o4h,Block.2021}. In contrast, here we report on the electronic thermal transport from the heat of a single photon at low temperatures.

We will study the performance of graphene single-photon bolometers resulting from the dissipative thermal diffusion towards the superconducting readout. In Sec.\ \ref{sec:1st}, we will introduce the differential equations that govern such processes, the initial and boundary conditions, and the different e-ph coupling regimes in graphene. In Sec.\ \ref{sec:d2}, we will solve the one-dimensional (1D) problem using an exact analytical solution for a toy model, before turning to numerical simulations in Sec.\ \ref{sec:d34}. Then, we will calculate the performance of graphene bolometers (Sec.\ \ref{sec:performance}) and extend our model to quasi-1D to account for the finite width of a graphene absorber (Sec.\ \ref{sec:quasi1d}). We conclude with an outlook envisioning a spatial-mode-resolving detector based on these results. 

\section{Dissipative heat diffusion of graphene electrons}\label{sec:1st}

We begin by considering the dissipative heat diffusion model depicted in Fig.\ \ref{fig:concept}, with the absorption of a photon into a graphene sheet with a superconducting readout nearby. We take the incoming light to be focused to a diffraction-limited spot with its size on the order of the photon wavelength, $\sim\lambda$. The absorption efficiency of graphene is only $\sim$2\% for light at normal incidence\cite{Sarma.2011}, but this can be enhanced with distributed Bragg reflectors\cite{Vasić.2014}, evanescent coupling to a proximitized waveguide \cite{Wang.2020waveguide}, or a nano-photonic cavity\cite{Efetov.2018}. For longer wavelength photons, $\lambda > 10~\mu$m, an antenna can guide and impedance match the incident radiation to the graphene\cite{Castilla.2019,Lee.2020}. 

\begin{figure}[t]\centering
\includegraphics[width=\columnwidth]{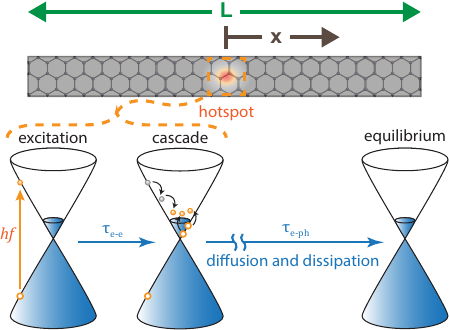}
\caption{\textbf{Photoexcitation cascade and timescales for diffusion and dissipation of electronic heat.} The interband excitation from the absorption of an infrared photon with energy $hf$ leads to a cascade of multiple electron-electron scattering events over a time $\tau_{\text{ee}}\sim100$ fs. This generates a local hotspot of radius $\xi$, the initial condition in our model. The timescale over which electrons thermalize to the base temperature via e-ph scattering, $\tau_{\text{e-ph}}$, can be several orders of magnitude longer. During this time, heat spreads through the graphene by electron diffusion and dissipates into the lattice by e-ph coupling. The rise and fall of the electronic temperature at a distance $x$ away from the hotspot can be measured in a graphene sample of length $L$.}
\label{fig:cascade+timescale}
\end{figure}

Microscopically, the absorption of a photon in graphene is mediated by the interband excitation of a valence band electron\cite{Massicotte.2021} (see Fig.\ \ref{fig:cascade+timescale}). The excited electron produces a localized hotspot via multiple e-e scatterings on a fast timescale\cite{Brida.2013,Johannsen.2013,Somphonsane.2013,Gallagher.2019,Block.2021}, $\tau_{\text{ee}}$, pre-empting the loss of energy to optical phonons and creating a locally thermalized distribution of charge carriers with a hotspot temperature, $\Theta_{\text{hot}}$. This hotspot diffuses out through the length of the graphene by random thermal motion to reach the superconducting sensor. Meanwhile, the heat is also lost to the phonon bath on a timescale $\tau_{\text{e-ph}}\gg\tau_{\text{ee}}$. Our model only considers timescales much longer than $\tau_{\text{ee}}$.

\subsection{Electronic thermal diffusion and dissipation}
We establish the differential equation including diffusion and dissipation to describe the thermal propagation of electrons in graphene. In a small area $\mathcal{A}$, heat can either diffuse out or dissipate to the lattice via e-ph coupling according to:
\ba c_e\frac{\partial\Theta_e}{\partial t} = \nabla\cdot\left(\kappa_e\nabla \Theta_e\right) -\Sigma(\Theta^\delta_e-\Theta^\delta_0). \label{eqn:heatdiffeqn} \ea 
The left-hand side of Eq.\ \ref{eqn:heatdiffeqn} is the rate of change of the internal energy of the electrons, with $c_{e}$ and $\Theta_{e}$ being the electronic specific heat per unit area and the electronic temperature, respectively. In monolayer graphene, the Fermi energy $E_{F}= \hbar v_{F} k_{F}$ has a Dirac-like dispersion relation, where $\hbar = h/2\pi$ is the reduced Planck's constant, $v_F=10^6$ m/s is the Fermi velocity, and $k_F = \sqrt{\pi n}$ is the Fermi wavevector with $n$ the charge carrier density. For typical 2D charge carrier densities of $10^{10}-10^{12}$ cm$^{-2}$, the Fermi temperature, $T_F = E_F/k_B$, ranges from 135 to 1350 K, far higher than typical bolometer operating temperatures of $\sim$1 K. In this degenerate regime the electronic specific heat is a linear function of temperature, $c_e = \gamma \Theta_e$, where $\gamma=4\pi^{5/2}k_B^2 n^{1/2}/(3 h v_F)$ is the Sommerfeld coefficient. The corresponding density of states is $g(E_{F})=2|E_{F}|A/(\pi \hbar^2 v_F^2)$, growing linearly with $E_F$ measured from the charge neutrality point\cite{Sarma.2011}. This contrasts with the constant density of states in two-dimensional (2D) electron gases with parabolic bands, and leads to an extraordinarily small specific heat $c_e \sim 1 ~k_B/\mu$m$^2$ at 20 mK. 

The first term on the right-hand side of Eq.\ \ref{eqn:heatdiffeqn} is the divergence of the heat flow into or out of $\mathcal{A}$. By Fourier law, the heat flow is $-\kappa\nabla\Theta_e$ with $\kappa_e$ the electronic thermal conductivity. Per the Wiedemann-Franz relation, $\kappa_e = \sigma_e\mathcal{L}_0\Theta_e$, where $\sigma_e$ is the electrical conductivity and $\mathcal{L}_0 = \pi^2k_B^2/3e^2$ is the Lorenz number with $e$ the electron charge.

The second term on the right-hand side of Eq.\ \ref{eqn:heatdiffeqn} describes the heat transfer per unit area between electrons and the lattice via e-ph coupling\cite{Viljas.2010,Chen.2012,Massicotte.2021}. Similar to Stefan-Boltzmann blackbody radiation, it is a high power law in temperature, originating from the integral of bosonic and fermionic occupations and densities of states in a Fermi golden-rule calculation. Here $\Sigma$ is the e-ph coupling parameter and $\delta$ is the constant exponent of a temperature power law that depends on sample quality.

In Eq.\ \ref{eqn:heatdiffeqn}, we have omitted the heat flow from the thermoelectric effect\cite{Massicotte.2021}. This simplification is justified by the negligible magnitude of the Seebeck coefficient at low temperatures according to the Mott relation\cite{Vera-Marun.2016}. At room temperatures, however, the thermoelectric effect is a sizable effect which can be used as a readout for bolometer or photodetector\cite{Gabor.2011,Cai.2014,Skoblin.2018}.

\subsection{Three regimes of e-ph coupling}

A key advantage of graphene is the weak e-ph coupling in Eq.\ \ref{eqn:heatdiffeqn}, which reduces heat dissipation to maintain a high $\Theta_e$ for long enough time to enable a high signal-to-noise of $\Theta_e$ readout. The strength of the e-ph coupling, i.e. both $\Sigma$ and $\delta$, can depend on the temperature and cleanliness of the graphene flake. Here, we only consider the e-ph (e-ph) coupling to acoustic phonons because optical phonons in graphene have energies $\sim$0.2 eV, well above the bolometer operating temperatures\cite{Sarma.2011,Song.201102t}. For the same reason, we will ignore the cooling channel by the surface phonons\cite{Rengel.2014,Crossno.2015,Massicotte.2021} as well.

The character of e-ph coupling differs when the electronic temperature is above or below the Bloch-Gr{\"u}neisen temperature, $\Theta_{\text{BG}} = 2 \hbar s k_{F} / k_B$, where $s$ is the sound velocity in graphene\cite{Viljas.2010}. For $T<\Theta_{\text{BG}}$, the average phonon momentum is smaller than the size of the Fermi surface, so the accessible momentum space for e-ph scattering is greatly reduced by the need to conserve momentum\cite{stormer.1990,Viljas.2010,Efetov.2010}. As a result, heat deposited in electronic degrees of freedom can remain trapped for an extended period of time, propagating efficiently across $\mu$m-scale devices with little loss to the lattice. At the carrier densities $n \sim 10^{12} \text{cm}^{-2}$ considered in this work, $\Theta_{BG} \approx 100$ K. Therefore, we will only consider the e-ph coupling below $\Theta_{\text{BG}}$, and in three regimes of sample quality: 1) the clean limit, where impurities are scarce and e-ph coupling is weak, 2) the supercollision regime, where disorder enhances e-ph coupling, and 3) resonant scattering, where edge defects enhance e-ph coupling.

The clean limit can be experimentally achieved in pristine, hexagonal boron nitride (hBN) encapsulated graphene at low temperatures. In these devices the mean-free-path, $l_{mfp} = \hbar \mu k_{F} / e$, of charge carriers is longer than the typical inverse phonon momentum. This limit is characterized by a $\delta$ = 4 power law dependence and $\Sigma=\pi^{5/2} k_{B}^{4} D^{2} n^{1/2}/(15 \rho_{m} \hbar^{4} v_{F}^{2} s^{3})$, with $D \simeq 18$~eV being the deformation potential and $\rho_{m}=7.4 \times 10^{-19}$ kg $\mu$ m$^{-2}$ the mass density of graphene. In real devices, this regime is often reported when the perimeter to area ratio of the graphene flake is small\cite{Fong.2012, Betz.2012, Draelos.2019o4h, McKitterick.2016}.

Disorder can enhance the e-ph coupling by relaxing the constraints on momentum conservation in a process known as supercollision cooling\cite{Chen.2012,Song.2012}. This regime is characterized by a disorder temperature, $\Theta_{\text{dis}} = h s/k_{B} l_{mfp}$. When $\Theta_e < \Theta_{\text{dis}}$, disorder relaxes the momentum conservation requirement by mediating the scattering process. This enhances the heat transfer between electrons and phonons by a transition to $\delta = 3$ power law and $\Sigma=2\zeta(3)k_B^3D^2n^{1/2}/(\pi^{3/2}\rho_m\hbar^3v_F^2s^2l_{mfp})$, with $\zeta$ the Riemann zeta function\cite{Chen.2012}. In this regime, typically realized in dirty, un-encapsulated graphene, hot charge carriers will more rapidly thermalize with the lattice\cite{Song.2012,Chen.2012,Graham.2013,Betz.2013,Fong.2013,Ma.2014}.

Scanning temperature probes have revealed a role for a third regime of resonant scattering due to the trapping of hot charge carriers on atomic defects at the edges of the graphene flake\cite{Wehling.2010,Halbertal.2017,Kong.2018}. The scattering is mostly localized at graphene edges in high quality encapsulated samples, suggesting device fabrication may play a role\cite{Halbertal.2017,Draelos.2019,Lee.2020}. Similar to supercollision cooling, the result is an enhancement in e-ph coupling. In this regime, $\delta$ is predicted to be either 3 or 5 \cite{Kong.2018}. 

For our numerical calculations we employ typical values of numerous graphene material parameters (see Tab.\ \ref{tab:parameters}) and use experimentally reported values of $\delta$ and $\Sigma$ (see Tab.\ \ref{tab:regimes}), while noting more work is needed to confirm the power law and $\Sigma$ in the resonant-scattering regime for different fabrication methods. In contrast to the clean limit, a power law scaling of $\delta =3$---which may correspond to either the supercollision or the resonant scattering regimes---is typically reported in devices where the graphene is not encapsulated in hBN or has a large perimeter to area ratio\cite{Betz.2012,Graham.2013, Somphonsane.2013, Eless.2013,Fong.2013,Draelos.2019, Lee.2020}.

\begin{table}[t]
\centering
\begin{tabularx}{\columnwidth} { 
  >{\raggedright\arraybackslash}X 
  >{\hsize=.2\hsize\arraybackslash\centering}X 
  >{\hsize=\hsize\raggedleft\arraybackslash}X }
\hline\hline
\multicolumn{3}{c}{Graphene parameters in numerical calculations}\\
\hline
Electron density & $n_0$ & \rule{0pt}{1em}$2.0 \times 10^{12} \; \text{cm}^{-2}$\\
Electron mobility & $\mu$ & $1.0 \times 10^{5} \; \text{cm}^{2} \text{V}^{-1} \text{s}^{-1}$\\
Electron mean free path & $l_{mfp}$ & $1.7 \; \mu$m \\
Graphene speed of sound & $s$ & $2.6 \times 10^{4} \; \text{m s}^{-1}$\\
MLG deformation potential & $D$ & $2.9 \times 10^{-18} \; \text{J}$\\
MLG mass density & $\rho_{M}$ & $7.4 \times 10^{-19}$ kg~$\mu$m$^{-2}$\\
Fermi velocity & $v_{F}$ & $1.0 \times 10^{6} \; \text{m s}^{-1}$\\
Electrical conductivity & $\sigma$ & 0.032 S \\
Sommerfeld coefficient & $\gamma$ & $9.5 \times 10^{-10}$ W s m$^{-2}$K$^{-2}$ \\
Diffusion constant & $\mathcal{D}$ & 0.82 m$^2$ s$^{-1}$ \\
Photon frequency & $f$ & 193.4 THz\\
Photon wavelength & $\lambda$ & $1550 \; \text{nm}$\\
Base temperature & $\Theta_0$ & 0.020 K\\
Disorder temperature & $\Theta_{dis}$ & 0.73 K \\
\hline
\end{tabularx}
\caption{\textbf{List of graphene parameter to model single-photon bolometers in this report.} These numerical parameters are taken from various experimental reports\cite{Fong.2013,McKitterick.2016}.}
\label{tab:parameters}
\end{table}

\subsection{Boundary and Initial conditions}\label{sec:bic}
We will use the boundary condition of zero heat flow because the heat diffusion at graphene-superconductor interfaces at temperatures well below the superconducting transition temperature can be substantially suppressed, as Cooper pairs carry no thermodynamic entropy and cannot conduct heat\cite{Satterthwaite.1962,Fong.2013}. Thanks to both theoretical \cite{Song.2013ykg} and experimental \cite{Brida.2013,Johannsen.2013,Ulstrup,Ruzicka.2010,Gallagher.2019,Block.2021} efforts, the initial temperature profile of the hotspot after the photon absorption is reasonably understood and shall be modeled as a Gaussian with a half-width-half-maximum, $\xi$:
\ba \Theta_e(t=0) &=& \Theta_{\text{hot}} e^{-(x-x_0)^2/2\xi^2} + \Theta_0,\label{eqn:ic}\ea
where $x_0$ is the location of the photon absorption, and $\Theta_{\text{hot}}$ and $\Theta_0$ are the peak hotspot and base temperature, respectively. It will be useful for our analysis to have an initial condition which is Gaussian in $\Theta_e^2$ instead of $\Theta_e$:
\ba \Theta_e^2(t=0) &=& \Theta_{\text{hot}}^2 e^{-(x-x_0)^2/\xi^2}+\Theta_0^2. \label{eqn:ic2} \ea
For $\Theta_{\text{hot}} \gg \Theta_0$, Eqns.\ \ref{eqn:ic} and \ref{eqn:ic2} are approximately equivalent. The average energy of electrons in the hotspot after the initial photo-excitation cascade\cite{Song.201102t,Tielrooij.2013} will determine $\Theta_{\text{hot}}$. For near-IR photons, this temperature is typically on the scale of 100s to 1000s of Kelvin, which is $\gg\Theta_0$, justifying Eqn.\ \ref{eqn:ic2}. To estimate $\Theta_{\text{hot}}$, we equate the photon energy with the integrated heat capacity with respect to temperature, yielding\cite{Walsh.2017}: 
\ba hf &=& \int_{\Theta_0}^{\Theta_{\text{hot}}} \pi\xi^2\gamma\Theta_e \, d\Theta_e \label{eqn:hotspot1} \\
 \Theta_{\text{hot}}^2 &=& \frac{2hf}{\pi\xi^2\gamma} + \Theta_0^2 \label{eqn:hotspot2}.\ea
$\Theta_{\text{hot}}=$ 100~K corresponds to $\xi = 94$ nm, which is consistent with values previously reported in the literature\cite{Song.2013,Block.2021}. Since $\xi$ is much smaller than the size of the graphene flake, the initial heating profile is effectively a point source in our model.

\begin{table}[t]
\centering
\begin{tabularx}{\columnwidth} { 
  >{\raggedright\arraybackslash}X 
  >{\hsize=.5\hsize\arraybackslash\centering}X 
  >{\hsize=2\hsize\arraybackslash\centering}X 
  >{\hsize=.5\hsize\arraybackslash\centering}X 
  >{\raggedleft\arraybackslash}X }
\hline
\hline
Scattering & $\delta$ & $\Sigma$ & $\tau_{e-ph}$ & $l_{th}$ \\
\hline
fictitious & 2 & - & - & - \\
resonant\cite{Lee.2020,Draelos.2019} & 3 & 1 Wm$^{-2}$K$^{-3}$ & 16 ns & 114 $\mu$m \\
supercollision & 3 & $1.38\times 10^{-3}$ Wm$^{-2}$K$^{-3}$ & 11 $\mu$s & 3.1 mm \\
clean limit & 4 & $3.10\times 10^{-2}$ Wm$^{-2}$K$^{-4}$ & 19 $\mu$s & 4 mm \\
\hline
\end{tabularx}
\caption{\textbf{Different regimes of e-ph coupling in graphene.} The $\delta = 2$ case, though fictitious, has an analytical solution to Eqn. \ref{eqn:heatdiffeqn2}, making it a useful tool for developing intuition and checking numerical calculations.Numerical parameters for resonant scattering regime are estimated from experiments\cite{Lee.2020,Draelos.2019}. Numerical parameters for supercollision regime and clean limit are calculated using the parameters in Table \ref{tab:parameters} and formulas for $\Sigma$ in Section \ref{sec:1st}.B}
\label{tab:regimes}
\end{table}

\subsection{Simplifying to one Dimension}
As we are interested in how to keep the photon input far enough from the superconducting electrodes of the readout circuit to avoid Cooper pair breaking, the optimal aspect ratio of our graphene bolometer would be a long strip as depicted in Fig.\ \ref{fig:cascade+timescale} to minimize the total heat capacity. Therefore, we shall simplify our dissipative heat diffusion differential equation, Eqn. \ref{eqn:heatdiffeqn}, to 1D:
\ba \gamma \Theta_{e} \frac{\partial \Theta_{e}}{\partial t} = \frac{\partial}{\partial x} \left( \sigma \mathcal{L}_{0} \Theta_e \frac{\partial}{\partial x} \Theta_e \right) - \Sigma \left( \Theta_{e}^{\delta} - \Theta_0^{\delta} \right) \label{eqn:heatdiffeqn_inbetween} \ea
using $c_{e} = \gamma \Theta_{e}$, and Wiedemann-Franz law. Furthermore, we can rewrite the equation in terms of $\Theta_{e}^2$:
\ba \frac{\partial}{\partial t} \Theta_e^2= \mathcal{D}\frac{\partial^2}{\partial x^2}\Theta_e^2 - \frac{2\Sigma}{\gamma}\left( \Theta_{e}^{\delta} - \Theta_0^{\delta} \right) \label{eqn:heatdiffeqn2}\ea
in which we define a diffusion constant, $\mathcal{D}$, such that
\ba \mathcal{D} &=& \frac{\sigma \mathcal{L}_0}{\gamma}\label{eqn:diffuse}\\
&=& \frac{E_F\mu_e}{2e}\label{eqn:einstein}.\ea
where $\mu_e$ is the electron mobility, $\mu_e=\sigma/ne$, and Eqn.\ \ref{eqn:einstein} is the Einstein relation\cite{Rengel.2013,Block.2021}. For $\mu_e \simeq 10^{5} \; \text{cm}^{2}\text{V}^{-1}\text{s}^{-1}$ and density $2.0 \times 10^{12} \; \text{cm}^{-2}$, the typical $\sigma$ value is 0.032~S, giving a numerical value of $\mathcal{D}$ of 0.82 $\text{m}^2\text{s}^{-1}$, which is consistent with measured values\cite{Ruzicka.2010,Block.2021}. In general, this non-linear diffusion equation has no analytical solution for arbitrary values of $\delta$, with the exception of some special cases.

\begin{figure*}\centering
\includegraphics[width=2\columnwidth]{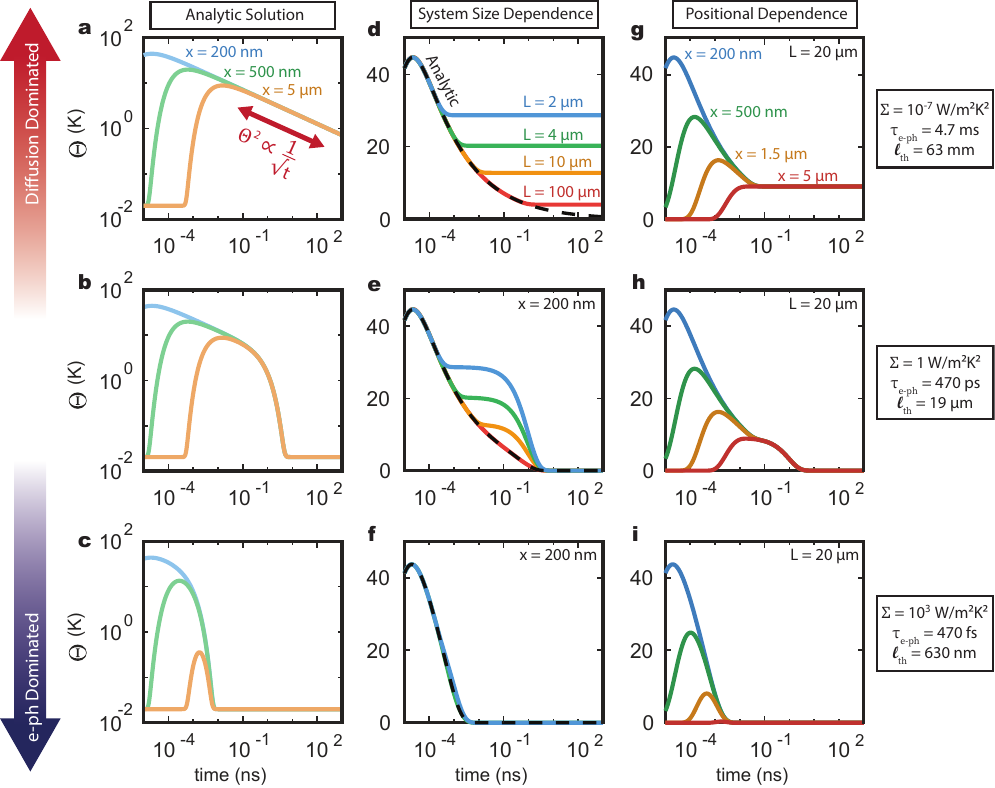}
\caption{\textbf{(a-c) Analytic solution for $\delta$ = 2 in an infinitely long sample} --- Results of numerical calculations of our analytic solution of the one-dimensional heat equation with an e-ph diffusion term. By adjusting the strength of the e-p coupling constant $\Sigma$, we can observe three regimes of heat diffusion in graphene. In Panel a, when the e-ph coupling is weak, the system is dominated by diffusion, characterized by the persistent $1/\sqrt{t}$ power law in $\Theta^2$. In Panel c, when the coupling is strong, the thermal propagation is dominated by dissipative e-ph effects. When the two effects are more closely balanced, like in Panel b, we can see a mix of diffusive and dissipative effects playing out on different timescales. \textbf{(d-f) System size dependence for $\delta$ = 2: analytic vs. numerical solution} --- Moving from infinitely long samples to graphene of finite length necessitates solving the heat equation numerically. Panels d-f compare the analytic solution (black dotted line) to the numerical results for various lengths of graphene. The numerical results line up well with the analytic solution for long samples. The plateau in temperature is explained by the entire sheet of graphene coming to a uniform temperature, which persists for a timescale determined by the e-ph coupling. \textbf{(g-i) Positional dependence in numerical solutions for $\delta$ = 2} --- Panels g-i show the simulated temperature response measured at various lengths away from the hotspot for a photon absorbed in the center of a 20 $\mu m$ piece of graphene. Predictably, the heat from the incoming photon takes longer to propagate to spots further away on the graphene. After approximately 100 picoseconds, all locations on the graphene reach a uniform temperature.}
\label{fig:grid3by3}
\end{figure*}

\section{Analytic description of the $\delta$ = 2 case} \label{sec:d2}
We can analytically solve Eqn. \ref{eqn:heatdiffeqn2} when $\delta = 2$ and the graphene sample is infinitely long. Using the initial condition Eqn. \ref{eqn:hotspot1} with $x_0=0$, the solution is:
\ba\Theta_e(x, t) = \left[\Theta_0^2+e^{-\frac{t}{\tau_{\text{e-ph}}}}e^{-\frac{(x-x_0)^2}{\xi^2+4\mathcal{D}t}}\cdot\frac{\Theta^2_{\rm{hot}}}{\sqrt{1+4\mathcal{D}t/\xi^2}}\right]^{1/2}\label{eqn:d2}\ea
where $\tau_{\text{e-ph}}^{(\delta=2)} = \gamma/2\Sigma$ is the thermal time constant bottlenecked by e-ph coupling when $\delta = 2$. While the first term, $\Theta_0^2$, comes from the base temperature, the second term describes the rise and fall of $\Theta_e$, within which the first exponential-decay factor is due to the e-ph coupling with time constant $\tau_{\text{e-ph}}$. When there are additional channels of heat dissipation, such as radiative coupling or contact to normal-metal electrodes, we can replace $\tau_{\text{e-ph}}$ with a more general thermal time constant, $\tau_{\text{th}}$. $\tau_{\text{th}} \leq \tau_{\text{e-ph}}$ because additional conductance channels will speed up the thermal dissipation. In the linear response regime, $\tau_{\text{th}}$ is given by the ratio of specific heat to the total thermal conductivity\cite{Walsh.2017}. The second factor provides the initial rise of temperature at position $x$ from the source at $x_0$ due to diffusion of heat, while the third factor describes the fall of $\Theta_e$ as heat diffuses further away.

Fig.~\ref{fig:grid3by3} plots Eqn.\ \ref{eqn:d2} with increasing $\Sigma$ from Fig.~\ref{fig:grid3by3}a to c. When diffusion dominates (Fig.~\ref{fig:grid3by3}a), the heat can diffuse throughout the sample. At a position $x$, the temperature will rise at $t\simeq (x-x_0)^2/\mathcal{D}$, and subsequently fall with a power law of $t^{-1/4}$. The latter is the expected diffusive behavior, i.e. $\Theta_e^2 \propto 1/\sqrt{t}$, in contrast to $\Theta_e \propto 1/\sqrt{t}$ if both $c_e$ and $\kappa_e$ do not depend on $\Theta_e$. As dissipation rises, we can still observe diffusive behavior when $t<\tau_{\text{e-ph}}$, but $\Theta_e$ falls at a faster rate than $t^{-1/4}$ (Fig.~\ref{fig:grid3by3}b) because of the high temperature power law of $\delta$. $\Theta_e$ decays at $\simeq\tau_{\text{e-ph}}^{(\delta=2)}$, independent of the position $x$. Deeper in the e-ph dominated regime, the decay timescale is fast enough to compete with the initial temperature rise from thermal diffusion (Fig.~\ref{fig:grid3by3}c). In this limit, the heat from the photon may not reach the sensing element of the detector before it is lost to the lattice.

\begin{figure}\centering
\includegraphics[width=0.9\columnwidth]{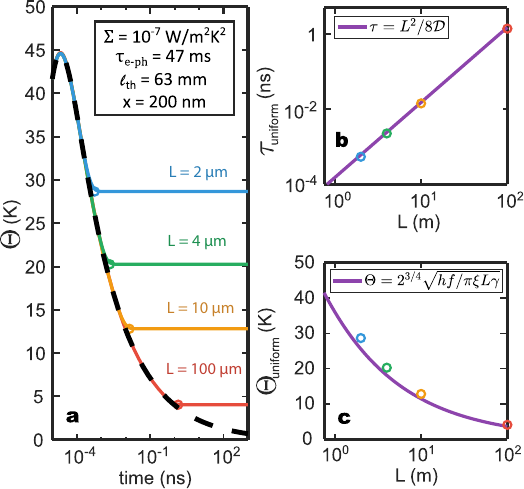}
\caption{\textbf{Diffusive behavior for $\delta=2$} --- Numerical calculations showing the details of the temperature plateau seen in the diffusive limit. \textbf{(a)} Equivalent to Fig.~\ref{fig:grid3by3}d, the behavior of different lengths of graphene in different colors (with the analytic solution for a sample of infinite length plotted as a dashed black line). \textbf{(b)} The timing of the plateau behavior $\tau_{\text{uniform}}$ plotted against sample length L, which follows $L^{2}/8\mathcal{D}$, characteristic of diffusion. \textbf{(c)} The temperature of the plateau $\Theta_{\text{uniform}}$ against system length, which is predicted by Eqn. \ref{eqn:uniform T}.}
\label{fig:delta2position}
\end{figure}

To investigate the thermal behavior of the system at finite size, we solve Eqn. \ref{eqn:heatdiffeqn} numerically for a graphene sheet of length $L$ with a hotspot in its middle. The second column of Fig. \ref{fig:grid3by3} plots these numeric solutions alongside the analytic solutions from the first column. Initially, all of these finite-system solutions follow the same initial rise and fall of $\Theta_e$ observed in the infinite-system analytic solutions. In the diffusive limit (Fig.~\ref{fig:grid3by3}d), the finite size solutions reach a plateau in temperature after a certain time, with shorter samples settling at a higher temperature than longer samples. From $\Theta_e$ at different positions along a $20~\mu$m sheet plotted in Fig.~\ref{fig:grid3by3}g, we confirm that $\Theta_e$ plateaus when the entire length of the graphene thermalizes to a uniform temperature, $\Theta_{\text{uniform}}$. Once the electron temperature is uniform, diffusion is complete and only dissipation can further lower the temperature. We can understand this behavior better by plotting the results in Fig. \ref{fig:delta2position} with the circles in the panel marking the time, $\tau_{\text{uniform}}$, at which graphene sheets of different lengths reach the plateau. Fig. \ref{fig:delta2position}b plots $\tau_{\text{uniform}}$ versus $L$. $\tau_{\text{uniform}}$ clearly follows the expected behavior for diffusion, $\tau_{\text{uniform}} = (L/2)^2/2\mathcal{D}$ (solid line), where $L/2$ is the distance from the hotspot to either end of the graphene flake. Evaluating Eqn. \ref{eqn:d2} at $t = L^2/8\mathcal{D}$, Fig. \ref{fig:delta2position}c shows $\Theta_{\text{uniform}}$ for different $L$. We can estimate $\Theta_{\text{uniform}}$ by considering our analytic solution in the diffusive limit. When $\xi^{2}/\mathcal{D} \ll t \ll \tau_{\text{e-ph}}$ and $\Theta_{\text{hot}} \gg \Theta_0$, Eqn. \ref{eqn:d2} simplifies to \ba \Theta_e	\simeq \frac{\Theta_{\text{hot}}}{\left(1+4\mathcal{D}t/\xi^{2} \right)^{1/4}}.\label{eqn:uniform T} \ea Substituting $\tau_{\text{uniform}}$ for t, and the estimation of $\Theta_{\text{hot}}$ from Eqn. \ref{eqn:hotspot2}, we have:  \ba \Theta_{\text{uniform}} \simeq 2^{3/4}\sqrt{\frac{hf}{\pi\xi L \gamma}}. \label{eqn:uniform T 2} \ea Comparison with Eqn.~\ref{eqn:hotspot2}, Eqn.~\ref{eqn:uniform T 2} suggests that the factor, $\pi\xi L$, is the effective area of the 1D graphene flake, with $\Theta_{\text{uniform}}$ playing the role of hot carrier temperature when the photon energy is distributed evenly across the entire graphene. Eqn. \ref{eqn:uniform T 2} (solid line in Fig.~\ref{fig:delta2position}c) well approximates $\Theta_{\text{uniform}}$ from our numerical calculation.

In Fig. \ref{fig:grid3by3}e, we move beyond the diffusive limit. Here we can still see the plateau behavior, but, similar to Fig. \ref{fig:grid3by3}b, the temperature subsides quickly after the timescale $\tau_{\text{e-ph}}^{(\delta=2)}$. Beyond a certain sample length, however, the timescale of e-ph coupling is faster than the time at which uniform temperature is reached, so the plateau does not appear. The relevant length scale for this behavior is the thermal-diffusion length $l_{\text{th}} = \sqrt{\mathcal{D}\tau_{\text{e-ph}}}$\cite{Song.201102t}. In Fig. \ref{fig:grid3by3}e, the 100 $\mu$m sample exceeds the $l_{\text{th}}$ value of 19 $\mu$m, and so exhibits no plateau behavior. In the e-ph dominated regime (Fig. \ref{fig:grid3by3}f), the timescale of thermal dissipation is so short that the plateau behavior is never observed for the sizes of the graphene that we consider in this manuscript.

Lastly, Fig. \ref{fig:grid3by3}g-i plot $\Theta_e$ at different positions for $L = 20~\mu$m. In the intermediate regime, $\Theta_e$ reaches $\Theta_{\text{uniform}}$ before relaxing back to $\Theta_0$, following the same curve governed by e-ph coupling regardless of the position. In the dissipative regime (Fig.~\ref{fig:grid3by3}i), a uniform elevated temperature is never reached, since the timescale of e-ph coupling $\tau_{\text{e-ph}}$ is faster than the time $L^{2}/4\mathcal{D}$ required for diffusion to bring the graphene to uniform temperature.

\section{Solution for $\delta$ = 3 and 4}\label{sec:d34}

\subsection{Linearized differential equation}

Although there is no exact solution to the dissipative heat diffusion differential equation (Eqn.\ \ref{eqn:heatdiffeqn2}) for $\delta = 3$ or 4, we can gain more insight by linearization. Using the substitution of $u = \Theta^2$ to Taylor expand Eqn.\ \ref{eqn:heatdiffeqn2}, we find \ba \frac{\partial u}{\partial t} &=& \mathcal{D}\frac{\partial^2 u}{\partial x^2} - \frac{2\Sigma}{\gamma}\left(u^{\delta/2}-u_0^{\delta/2}\right)\label{eqn:Linearized1}\\
&\simeq& \mathcal{D}\frac{\partial^2 u}{\partial x^2} - \frac{\delta \Sigma u_0^{\delta/2 -1}}{\gamma}\left(\Delta u + ... \right)\label{eqn:Linearized2}\ea where $\Delta u = u - u_0$. While Eqn. \ref{eqn:Linearized1} is still exact, the linearization (Eqn.\ \ref{eqn:Linearized2}) contains an approximation for the dissipative term. Taking only the first term of the Taylor expansion, Eqn.\ \ref{eqn:Linearized2} becomes linear in $u$. As a result, the solution takes the exact same form as Eqn. \ref{eqn:d2}, but with a more general form for $\tau_{\text{e-ph}}:$  \ba \tau_{\text{e-ph}} = \frac{\gamma}{\delta \Sigma \Theta^{\delta - 2}} \label{eqn:tau} \ea As a result, most of the intuition built up in Sec. \ref{sec:d2} is applicable to the more realistic $\delta = 3$ and 4 cases, especially in the diffusion-dominated regime.

\begin{figure}\centering
\includegraphics[width=0.9\columnwidth]{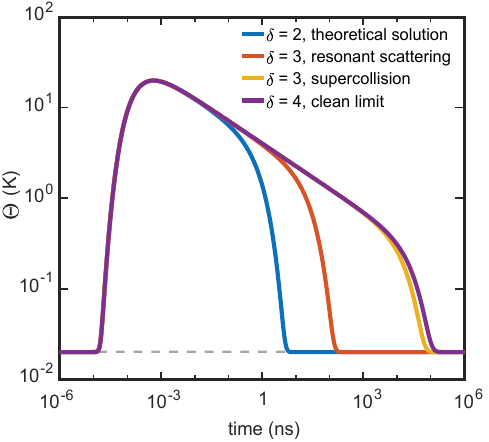}
\caption{\textbf{Analytic solution of the linearized differential equation for $\delta$ = 3 and 4 for an infinitely long sample} --- analytical calculations showing the different thermal responses of graphene following $\delta = 4$ vs $\delta = 3$ power laws, alongside a purely theoretical $\delta = 2$ solution. For these calculations, we set $\Sigma = 1$ for the $\delta = 2$ case, representing the regime in between diffusive and dissipative. The values of $\Sigma$ for the $\delta = 3$ and $\delta = 4$ cases are shown in Tab. \ref{tab:regimes}. Temperatures are measured at a distance of $x = 1~\mu$m away from the hotspot.}
\label{fig:delta34linearTime}
\end{figure}

We plot the linearized analytic solutions for each regime in Fig. \ref{fig:delta34linearTime}. The timing and magnitude of the initial peak in temperature track closely, i.e.~$t \ll \tau_{\text{e-ph}}$, across all e-ph regimes because the diffusion term in Eqn. \ref{eqn:Linearized2} is independent of $\delta$ and $\Sigma$. On the other hand, $\Theta_e$ drops differently depending on e-ph regimes with time constant given by Eqn. \ref{eqn:tau}. The linearized analytic solutions are ordered from fastest to slowest e-ph dissipation time according to their power law, with $\delta = 2$ being the most dissipative and $\delta = 4$ being the least. However, the magnitude of $\Sigma$ also plays a considerable role in determining the behavior of each regime: for instance, there is a greater difference between the two $\delta = 3$ regimes than between $\delta = 3$ and $\delta = 4$. Specifically, we have used a relatively clean graphene, with an electron mobility of $10^5$~cm$^2$/Vs and $l_{mfp}$ of 1.7~$\mu$m, even in the supercollision regime.

\subsection{Numerical solutions}

\begin{figure}\centering
\includegraphics[width=0.9\columnwidth]{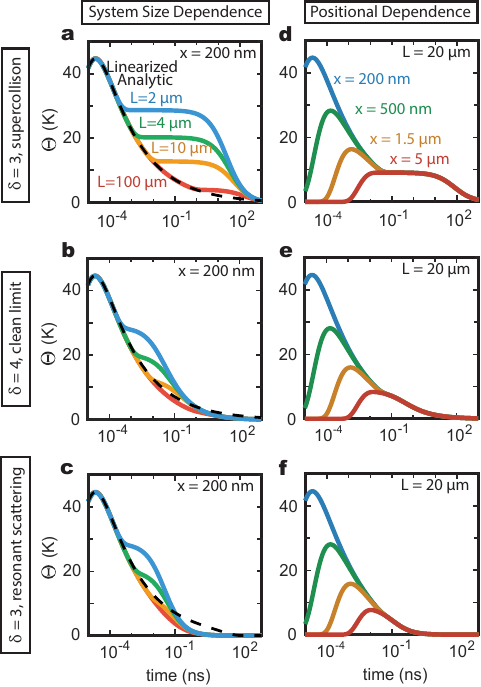}
\caption{\textbf{The size dependence of the numerical solution for $\delta$ = 3 and 4, for a finite-sized sample.} \textbf{(a-c)} The numerical simulation results for graphene samples of a variety of different finite sizes for each of the higher power-law e-ph coupling regimes, in order from most (Panel a) to least (Panel c) diffusive. The temperature response is measured 200 nm away from the hotspot. The linearized analytic solution (Eqn.\ \ref{eqn:d2} with $\tau_{\text{e-ph}}$ given by Eqn.\ \ref{eqn:tau}) is plotted as a black dashed line. Notably, the more diffusive the system, the better the linearized approximation performs. \textbf{(d-f)} The positional dependence of the same three regimes for a 20 $\mu$m long flake of graphene. The resonant scattering regime exhibits the fastest temperature decay, while the supercollision regime exhibits the slowest.}
\label{fig:delta34size}
\end{figure}

Similar to Fig.\ \ref{fig:grid3by3}d-i, Fig.\ \ref{fig:delta34size} shows the numerical solutions to Eqn. \ref{eqn:heatdiffeqn2} for a finite-sized graphene. In the supercollision regime (Fig.~\ref{fig:delta34size}a), $\Theta_e$ behaves similarly to the diffusive limit of $\delta = 2$, where $\Theta_e$ plateaus as the entire graphene reaches a uniform temperature, and $\Theta_{\text{uniform}}$ and $\tau_{\text{uniform}}$ depend on the system size in the same manner as shown in Fig.\ \ref{fig:delta2position}. In contrast, the shorter or vanishing plateaus of the clean and resonant scattering regimes imply a weaker diffusive behavior (Fig.\ \ref{fig:delta34size}b and c), leading to a drop to the baseline temperature faster than the linearized analytic solution. We emphasize that the qualitative behavior of $\Theta_e$ depends on both the numerical values of $\Sigma$ and $\delta$, rather than just the assignment of an e-ph scattering regime, as elaborated below.

Contrary to the behavior of the linearized analytic solution plotted in Fig.\ \ref{fig:delta34linearTime}, the $\delta = 3$ supercollision scattering mechanism exhibits much more diffusive behavior than the $\delta = 4$ clean limit when using the same numerical parameters. This is because $\Theta_e\gg\Theta_0$ in these calculations, well outside the linear response. The higher power law dissipation can relax $\Theta_e$ with a timescale, $\tau_{\text{relax}}$, which depends on both $\delta$ and $\Sigma$. We expect 
$\tau_{\text{relax}} < \tau_{\text{th}} \leq \tau_{\text{e-ph}} \label{eqn:tau relax}$. As illustrated in Fig.\ \ref{fig:delta34linearTime}, the value of $\tau_{\text{e-ph}}$ for the supercollision regime is slightly shorter than for the clean regime given the experimental conditions and parameters in Tab.\ \ref{tab:parameters}. But the lower power law means that $\tau_{\text{relax}}$ for the $\delta = 3$ supercollision regime is longer than the $\delta = 4$ clean regime when $\Theta_e\gg\Theta_0$. This is the reason for the stronger diffusive behavior observed in the supercollision regime (Fig. \ref{fig:delta34size}a and d) compared to the clean regime (Fig.\ \ref{fig:delta34size}b and e). On the other hand, the numerical value of $\Sigma$ is several orders of magnitude larger for the resonant scattering regime (see Tab. \ref{tab:regimes}), so that it exhibits the least diffusive behavior of the three regimes, despite having both a lower power law than the clean regime and a shorter $\tau_{\text{relax}}$. This result demonstrates the need to numerically solve the dissipative heat diffusion differential equation because qualitative details can depend on the experimental values of both $\Sigma$ and $\delta$. The same intuition for the faster-than-exponential dissipation for high power laws in the e-ph coupling can also explain why the numerical solutions drop below the analytic solution in linear response (black dotted line in Fig.\ \ref{fig:delta34size}).

The position dependence of $\Theta_e$ for $\delta = 3$ and 4 (Fig.~\ref{fig:delta34size}d-f) behaves similarly to the $\delta = 2$ solutions. For the supercollision regime, Fig.\ \ref{fig:delta34size}d plots the different locations on a graphene flake reaching a $\Theta_{\text{uniform}}$, maintaining it for a time, and eventually subsiding to $\Theta_0$. $\Theta_e$ in the clean and resonant scattering regimes (Fig.\ \ref{fig:delta34size}e and f) can also reach the same $\Theta_{\text{uniform}}$ as in Fig.~\ref{fig:delta34size}d, although the dissipation has already kicked in by the time $\Theta_e$ reaches $\Theta_{\text{uniform}}$. A subtle but important difference between Fig.\ \ref{fig:delta34size}e and f is the speed at which the sample thermalizes to $\Theta_0$ after $t\simeq \tau_{\text{uniform}}$. The tail of the temperature curve decays with a considerably gentler slope in Fig. \ref{fig:delta34size}e than in Fig. \ref{fig:delta34size}f, corresponding to several additional nanoseconds at elevated temperature for device in clean regime than for a device in resonant scattering one. This difference will prove to be important to the detector efficiency.

\begin{figure}[h]\centering
\includegraphics[width=0.9\columnwidth]{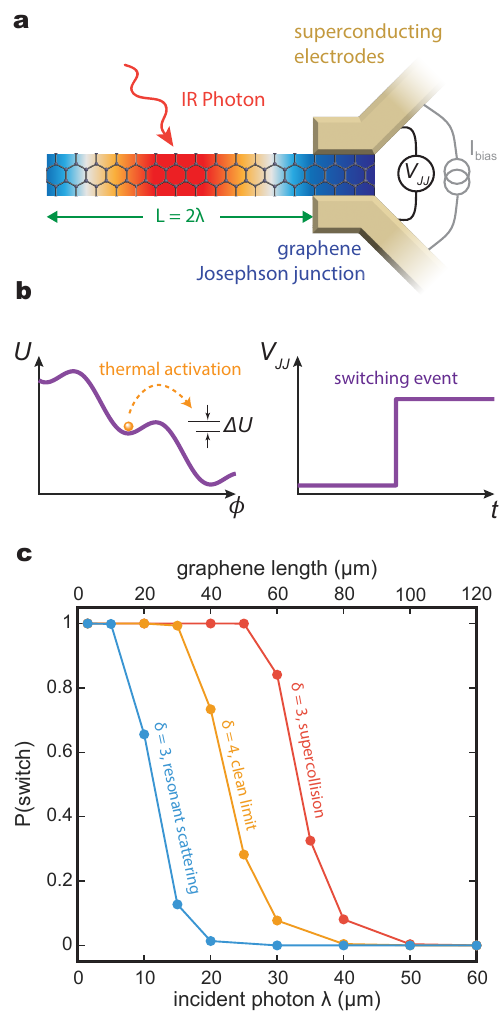}
\caption{\textbf{Josephson Junction bolometer performance.} \textbf{(a)} A schematic representing a Josephson junction-based bolometer with superconducting readout. The length of the device is set to be the minimum two times the photon wavelength, limited by diffraction. \textbf{(b)} The detection mechanism, whereby thermal fluctuations allow electrons to overcome the washboard potential, leading to thermally activated switching of the Josephson junction detectable via voltage probe.  \textbf{(c)} The simulated detection efficiency of such a device for the three e-ph coupling regimes, ignoring any geometric losses that might occur from various coupling mechanisms.}
\label{fig:efficiency}
\end{figure}

\section{Bolometer performance limited by dissipative heat diffusion}\label{sec:performance}

\subsection{Detector efficiency in the mid- and near-infrared regime}

We can now use the calculated $\Theta_e(t)$ at the sensor position, of distance $d$, from the photon input to study the performance of the graphene bolometer. To begin, we need to decide the heat sensor that shall be used because its operation mechanism can determine the overall sensitivity, efficiency, readout bandwidth, dark count, and timing jitter. For example, when applying resonator techniques to directly measure the $\Theta_e$in a normal-metal-insulator-superconductor\cite{Gasparinetti.2015} or Josephson junction\cite{Zgirski.2018}, we expect the bolometer sensitivity to depend on the noise temperature of the amplifier chain of the entire measurement system. On the other hand, when using a Josephson junction to indirectly detect the rise of $\Theta_e$ from the switching rate\cite{Walsh.2017,Walsh.2021}, one may evade the amplifier noise when the Josephson junction switches discontinuously from the superconducting to resistive state in response to the arrival of a photon. Such a DC measurement is straightforward to implement and we will use it to analyze the performance of a single-photon bolometer in this manuscript. In comparison, the kinetic inductance measurement\cite{Giazotto.2008} with a resonator is a faster and probably higher-sensitivity readout to register the fleeting rise and fall of $\Theta_e(d, t)$ due to absorbing a single photon. We will defer analysis of the resonator readout to a future work due to additional complexity arising from a $\tau_{\text{th}}$ that is shorter than the response time of the readout resonator.

Let us now consider the bolometer design. The first criteria is to keep the superconducting readout circuitry and the location of the photon input separated by a distance shorter than the thermal diffusion length, $l_{\text{th}}$. This requirement can easily be satisfied because $l_{\text{th}}$ is on the order of tens of micrometers, much longer than the photon wavelength. Since the temperature rise from photons scales inversely to the heat capacity, the graphene bolometer must be short in the transverse direction to minimize the total area, while remaining long enough in the longitudinal direction to maintain the separation between the photon input and superconducting readout. Fig.~\ref{fig:efficiency}a illustrates the graphene bolometer that we have modeled to assess its detection efficiency in the near and mid-IR regime. The photon input, a focused, diffraction-limited light spot with beam waist $\sim\lambda$, is located in the middle of the flake while the sensor, a Josephson junction with a graphene weak link, is attached to the end of the flake. We will set the length of the graphene absorber to be 2$\lambda$ to allow for the efficient coupling of photons\cite{Efetov.2018,Ye.2021}. As the photon wavelength increases from near-IR to mid-IR, we expect the detector efficiency to decrease. This is because the size of a diffraction-limited beam spot grows with increasing wavelength, requiring a larger graphene absorber. In a longer device, both the diffusive and dissipative parts of Eqn. \ref{eqn:d2} contribute to reduce $\Theta_e$ at the sensor position. Additionally, the lower photon energy creates a lower initial $\Theta_{\text{hot}}$, which can be modeled by lowering the photon frequency in Eqn. \ref{eqn:hotspot2} while keeping the hotspot size constant. Our goal in modeling this device is to understand the variation of detection efficiency across a wide range of the electromagnetic spectrum based on this simple design. For the detection of far IR photon, which is outside the scope of our discussion, we will need to switch from a long graphene absorber to an antenna-coupled device to focus the photon electric field closer to the readout element\cite{Castilla.2019}.

We model the graphene-based Josephson junction as a heat switch that changes state due to the heat from absorbing a single photon. This device concept\cite{Walsh.2017} can be understood phenomenologically through the resistively and capacitively shunted junction (RCSJ) model. This model describes Josephson junction dynamics in terms of a fictitious, massive phase particle residing inside the well of a tilted washboard potential which is dependent on the phase difference, $\phi$, between the two superconducting electrodes (Fig.\ \ref{fig:efficiency}b). The potential barrier, $\Delta U$, constraining the phase particle is given by: $\Delta U = 2 E_{J0}(\sqrt{1 - \gamma^2}-\gamma \, \cos^{-1}\gamma)$ where $E_{J0} = \hbar I_{c}/2e$ is the Josephson energy and $\gamma = I_b/I_c$, with $I_b$ the applied current bias and $I_c$ the critical current of the junction. While $E_{J0}$ sets the scale of the barrier height, $I_b$ can further tilt the washboard by increasing $\gamma$, hence lowering the potential barrier. When the phase particle rests inside the well, the device is in the superconducting state. If instead the phase particle is perturbed such that it falls down the well, i.e. with a finite $d\phi/dt$, it is in the normal state. When a photon is absorbed by the graphene, $\Theta_{e}$ rises. With a higher ratio of the thermal energy to the potential barrier, $k_B\Theta_{e}/\Delta U$, the phase particle can be excited out of the well, causing a transition to the normal state. This transition serves as a heat switch and will result in an increase in the voltage, $V_{JJ}$, measured across the junction (Fig. \ref{fig:efficiency}b), indicating the detection of a photon. The dark count, i.e. spontaneous switching without incident photons is predominately caused by macroscopic quantum tunneling\cite{Walsh.2017,Walsh.2021}.

\begin{table}
\centering
\begin{tabular}{ l c c } 
\hline\hline
\multicolumn{3}{c}{~~~~~~~~Josephson Junction Device Parameters~~~~~~~~}\\
\hline
Junction Capacitance & $C_{JJ}$ & 1.0 fF\\
Junction Resistance & $R$ & 50~$\Omega$\\
Critical Current & $I_{c}$ & 3.38 $\mu$A\\
Bias Current & $I_{\text{bias}}$ & 2.80 $\mu$A\\
Quality factor & Q & 0.11\\
Integration Time & $t_{\text{meas}}$ & $100$~ns\\
\hline
\end{tabular}
\caption{\textbf{Parameters of the Josephson Junction in our model.} These inputs were used along with a fit to an RCSJ model of a Josephson Junction to calculate the switching probability from the simulated spatiotemporal temperature response in graphene. These are typical numerical values from experiments\cite{Walsh.2017, Walsh.2021}.}
\label{tab:RCSJparameters}
\end{table}

Detection of the photon relies on understanding the switching rate of the junction from the superconducting to normal state. When the switching is thermally activated, its rate can be described as: \ba \Gamma = Ae^{-\Delta U/k_B\Theta_{e}}\label{eqn:SwitchingRate}\ea where $A$ is a proportionality factor given by: \ba A = \frac{\omega_{p}}{2\pi}\left(\sqrt{1 + \frac{1}{4Q^{2}}} - \frac{1}{2Q}\right)\ea with $Q = \omega_p R_n C_{JJ}$ being the quality factor of the harmonic well in the washboard potential with $R_n$ the junction's normal resistance, $C_{JJ}$ the junction's capacitance, and $\omega_p$ the plasma frequency of the Josephson junction. At a finite bias current, $\omega_p = \omega_{p0}(1-\gamma_{JJ}^2)^{\frac{1}{4}}$, where $\omega_{p0}= \sqrt{2eI_c/\hbar C_{JJ}}$ is the zero-bias plasma frequency. 

To calculate the probability of the switching of the Josephson junction from $\Gamma$ during the duration of the measurement, $t_{\text{meas}}$, we can use\cite{Walsh.2017}:
\begin{equation}
    P(\text{switch}) = 1 - e^{-\int_{0}^{t_{\text{meas}}} \Gamma dt}.
    \label{eqn:switching probability}
\end{equation} We numerically evaluate $P(\text{switch})$ using the experimental values from a typical graphene-based Josephson junction listed in Tab. \ref{tab:RCSJparameters} and plot the results in Fig. \ref{fig:efficiency}c. Regardless of the e-ph coupling regime, we can achieve a high, intrinsic efficiency for short wavelength photons because there is enough photon energy to diffuse to the sensor location though the short graphene absorber. However, as $\lambda$ lengthens, the efficiency falls as $\Theta_{\text{hot}}$ decreases. The integral in Eqn.\ \ref{eqn:switching probability} signifies that the importance of the temperature response is not only the height of the initial temperature peak, but also the elapsed time that the graphene device remains at elevated temperatures. As such, although the $\delta = 3$ resonant scattering and the $\delta = 4$ clean limit solutions in Fig. \ref{fig:delta34size} appear visually similar, the longer tail of the decaying temperature curve for $\delta = 4$ can significantly impact the thermally-activated switching rate, so a device exhibiting $\delta = 4$ behavior is better suited to detecting low energy photons. Predictably, the $\delta = 3$ supercollision case maintains the highest detector efficiencies at longer photon wavelengths due to its longer time at elevated $\Theta_e$. Similar to Fig.~\ref{fig:delta34size}, we note that the performance of a single-photon bolometer depends on both the numerical values of $\Sigma$ and $\delta$ in the e-ph coupling, rather than the simple assignment of e-ph regime. Specifically, we are using $\Sigma$ and $\delta$ from a relatively high-quality graphene in the supercollision regime that is achievable in experiments (Tab.~\ref{tab:parameters}).

These results inform the design of single-photon bolometers for different photon detection experiments. For photons with $\lambda < 5~\mu$m, a free-space-coupled single-photon detector should work well, as long as the device is appropriately short. For photons up to $15~\mu$m, the fabrication requirements become more stringent to remain in the supercollision regime. Finally, if trying to detect photons of $\lambda > 30~\mu$m, some compromises to the photon coupling must be made, or more sophisticated methods must be employed, in both the photon coupling design and device fabrication to maintain a high detection efficiency.

\subsection{Intrinsic limit of jitter due to spatial photonic modes}
\begin{figure}[t]
\includegraphics[width=0.9\columnwidth]{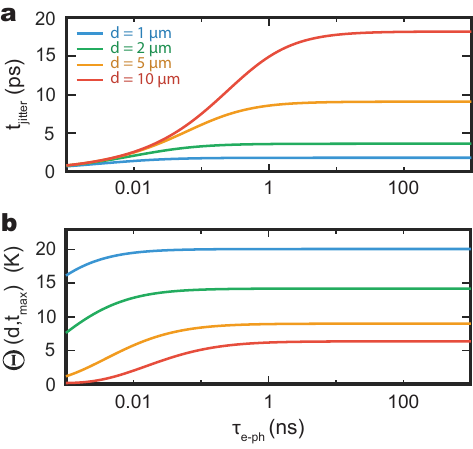}
\caption{\textbf{Dependence of timing jitter and maximum electron temperature on detector distance.} For a graphene single-photon bolometer with $\lambda$ = 1550 nm and $\mathcal{D}$ listed in Tab. \ref{tab:parameters}, \textbf{(a)} the timing jitter versus $\tau_{\text{e-ph}}$ for according to Eqn. \ref{eqn:jitter_corrected} and \textbf{(b)} the peak temperature in Panel a. In Panel b, we plot the electron temperature at the detector at the time of arrival $t_{max}$.}
\label{fig:jitter}
\end{figure}
In addition to detection efficiency, the timing jitter is another important parameter for sensitive photodetectors. The detector physics and process can impose the most fundamental limit to timing jitter\cite{Allmaras.2019}. We will apply our dissipative heat diffusion model to estimate this intrinsic limit when it is due to the finite size of the spatial mode of the photon input\cite{Korzh.2020}.

As described in Sec. \ref{sec:bic}, the hot spot initially created by absorbing the internal energy of a single photon is effectively a point source. The hot spot can be found anywhere within the beam spot with a probability distribution that follows from the intensity profile of the optical mode. This is because the rate of the photon absorption by graphene is proportional to the square of the electric field, and thus the spatial mode profile. For a diffraction-limited Gaussian beam, we expect the standard deviation of the variation in hot spot location to be on the order of $\lambda$.

To calculate time jitter, we use the peak time of $\Theta_e$, $t_{\text{max}}$ in the 1D model, as the time taken for the heat to diffuse from the hot spot to the sensor at a distance $d$ away from the focal point. In Eqn.~\ref{eqn:d2}, $\Theta_e$ hits the peak when: \ba t_{\text{max}} = \frac{\sqrt{\mathcal{D}\tau_{\text{e-ph}}\left(4(d-x_0)^2 + \mathcal{D}\tau_{\text{e-ph}}\right)} - \mathcal{D}\tau_{\text{e-ph}}}{4\mathcal{D}}. \label{eqn:peaktime_corrected} \ea To estimate the timing jitter, $t_{\text{jitter}}$, we shall take the difference of $t_{\text{max}}$ for $x_0 = \pm \lambda/2$, accounting for the location variation of the hot spot due to photon absorption, such that: \begin{widetext} \ba t_{\text{jitter}} = \frac{1}{4}\sqrt{\frac{\tau_{\text{e-ph}}}{\mathcal{D}}} \left(\sqrt{4(d+\lambda/2)^2 + \mathcal{D}\tau_{\text{e-ph}}} - \sqrt{4(d-\lambda/2)^2 + \mathcal{D}\tau_{\text{e-ph}}} \right) \label{eqn:jitter_corrected}\ea
\end{widetext}

Fig.~\ref{fig:jitter}a plots Eqn.~\ref{eqn:jitter_corrected} versus $\tau_{\text{e-ph}}$. In the diffusive limit, as $\tau_{\text{e-ph}} \rightarrow \infty$, $t_{\text{max}}$ is determined by the speed of the heat diffusion. Eqn.~\ref{eqn:peaktime_corrected} reduces to $t_{\text{max}} = (d - x_0)^2/2\mathcal{D}$ and the jitter approaches
\ba t_{\text{jitter}} = \frac{d\lambda}{\mathcal{D}}. \label{eqn:jitter}\ea
Intuitively, the dependence on $\lambda$ corresponds to a larger spatial uncertainty in photon absorption for longer wavelengths. A short distance, $d$, or faster diffusion, $\mathcal{D}$, can reduce the difference of the time taken for the heat to arrive at the sensor from different locations within the spatial mode of the photons. Using $\mathcal{D} = 0.82 $m$^2$s$^{-1}$ and taking $d \simeq \lambda$, $t_{\text{jitter}} = 2.7$~ps for 1500 nm wavelength IR photons. This value is comparable to SNSPDs\cite{Korzh.2020,Zadeh.2020} and can be improved by using higher-mobility graphene or hydrodynamic effect\cite{Lucas.201738q,Block.2021}. Paradoxically, Eqn. \ref{eqn:jitter_corrected} suggests that $t_{\text{jitter}}$ also improves at shorter $\tau_{\text{e-ph}}$. This is because the e-ph dissipation tends to reduce $t_{\text{max}}$ at the cost of a lower maximum electron temperature, $\Theta_e(x, t_{\text{max}})$ (Fig. \ref{fig:jitter}b). As a result, timing jitter and dissipation are at odds. As the efficiency of a single-photon bolometer relies heavily on a higher $\Theta_e(d, t)$ at the sensor component, it is detrimental to the detector efficiency to improve jitter by increasing the e-ph coupling. For better timing jitter and detector efficiency, the design principle is to operate in the diffusive limit while seeking to maximize $\mathcal{D}$ and minimize $d$.

\section{Quasi-one-dimensional model}\label{sec:quasi1d}
\begin{figure}[h]\centering
\includegraphics[width=0.9\columnwidth]{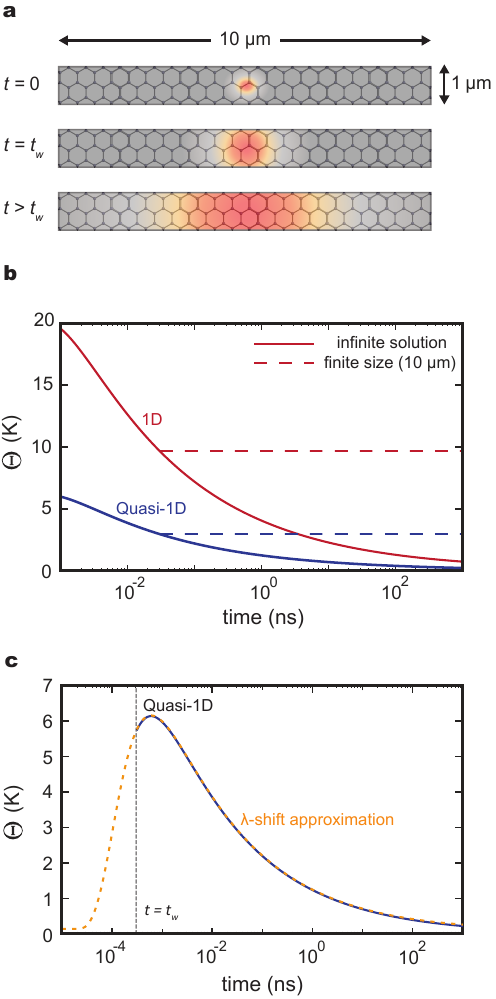}
\caption{\textbf{Quasi-one-dimensional model.} \textbf{(a)} The heat diffusion in a quasi-1D graphene system where the width is finite but much shorter than the length. Initially, the heat propagates out in 2 dimensions, but after a time $t_w$, the heat has propagated through the full width of the graphene and the system's behavior can be approximated by a 1D solution with a larger and cooler initial hotspot. \textbf{(b)} The different temperature profiles of the 1D vs. quasi-1D models for a graphene flake with a length of 10 $\mu$m and a width of 1 $\mu$m, measured at a distance of 1 $\mu$m from the center of the hotspot. \textbf{(c)} Comparison of the quasi-1D solution (black dashed line) with the 1D solution with a lower photon energy (yellow dashed line) as in Eqn. \ref{eqn:lambdashiftapprox}.}
\label{fig:2Dto1D}
\end{figure}

The previous sections consider the strictly 1D dissipative heat diffusion model with $\xi$ as the effective width suggested by Eqn. \ref{eqn:uniform T 2}. Here, we would like to understand how the finite width, $w$, in realistic samples affects our calculations and conclusions. Interestingly, the intuition that we built through the one-dimensional model will help us to approximate a quasi-one-dimensional solution for a narrow rectangular strip, i.e. $w \ll L$ (see Fig. \ref{fig:2Dto1D}).

We can extend our 1D initial condition in Eqn. \ref{eqn:ic2} to include a second spatial dimension in the transverse direction, $y$:
\ba \Theta_e^2(t=0)= \Theta_{\text{hot}}^2 e^{-\frac{(x-x_0)^2+(y-y_0)^2}{\xi^2}}+\Theta_0^2 = \Theta_{\text{hot}}^2 e^{\frac{-r^2}{\xi^2}}+\Theta_0^2.\label{eqn:2DinitialCondition}\ea with the center of the hotspot at $(x_0,y_0)$, and where the second equality presents the initial condition in polar coordinates, i.e.\ $r^2 = (x-x_0)^2+(y-y_0)^2$. The differential equation and solution of the dissipative heat diffusion in 2D is similar to that in 1D (Eqn.\ \ref{eqn:heatdiffeqn2} and \ref{eqn:d2}) and are given by:
\ba\frac{\partial}{\partial t} \Theta_e^2 &=& \mathcal{D}\left(\frac{\partial^2}{\partial r^2}+\frac{1}{r}\frac{\partial}{\partial r}\right)\Theta_e^2 - \frac{2\Sigma}{\gamma}\left( \Theta_{e}^{\delta} - \Theta_0^{\delta} \right) \label{eqn:heatdiffeqn_2d_a}\\
\Theta_e(r, t) &=& \left[\Theta_0^2+e^{-\frac{t}{\tau_{\text{e-ph}}}}e^{-\frac{r^2}{\xi^2+4\mathcal{D}t}}\frac{\Theta^2_{\rm{hot}}}{1+4\mathcal{D}t/\xi^2}\right]^{1/2}\label{eqn:d2_in_2D},\ea respectively. Qualitatively, our discussions and plots of the solution in 1D applies in 2D as well when the flake is infinitely large (Fig. \ref{fig:grid3by3}a-c and \ref{fig:delta34linearTime}) or circular in shape (Fig. \ref{fig:grid3by3}d-i, \ref{fig:delta2position}, and \ref{fig:delta34size}).

When the flake is a strip with $w\ll L$ (Fig. \ref{fig:2Dto1D}a), we can construct an approximate, quasi-1D solution. Here Eqn.\ \ref{eqn:d2_in_2D} is initially correct before the heat reaches the edge of the graphene flake in the $y$-direction. Similar to the finite width in 1D, after a characteristic time, $t_w$, we expect that $\Theta_{\text{2D}}(x=0,y,t)$ becomes uniform in and independent of the $y$ position at the center of the longitudinal direction, provided that $t_w \ll \tau_{\text{e-ph}}$. From our analysis and plots in Fig. 4, $t_w = (w/2)^2/\mathcal{D}$ such that Eqn. \ref{eqn:d2_in_2D} becomes:
\ba \Theta_{\text{2D}}(x, y, t_w) \simeq \left[\Theta_0^2 + e^{-\frac{t_w}{t_{\text{e-ph}}}}e^{-\frac{x^2+y^2}{\xi^2+w^2}}\frac{\Theta_{\text{hot}}^2}{1+(w/\xi)^2}\right]^{1/2}.\label{eqn:2Dto1Dhotspot}\ea 

We can now cast the problem back to 1D by applying Eqn.~\ref{eqn:2Dto1Dhotspot} with $y = 0$ as our new initial condition. Schematically shown as $t=t_w$ in Fig. \ref{fig:2Dto1D}a, the size of the hotspot at $t=t_w$ should have grown larger than $\xi$ in the new initial condition for our quasi-1D solution. Direct comparison of Eqn. \ref{eqn:2DinitialCondition} and Eqn. \ref{eqn:2Dto1Dhotspot} leads to the definition of the quasi-1D hotspot size, $\tilde{\xi}$, and temperature, $\tilde{\Theta}_{hot}$, as:
\ba \tilde{\xi} &=& \sqrt{\xi^2+w^2} \\
\tilde{\Theta}_{hot} &=& e^{-\frac{t_w}{2t_{\text{e-ph}}}} \left(\frac{\xi}{\tilde{\xi}}\right) \Theta_{\text{hot}}, \ea respectively. The new quasi-1D initial condition to capture the effect of the finite width of the device is:
\ba \tilde{\Theta}_e^2(x,\tilde{t}=0) = \tilde{\Theta}_{hot}^2 e^{\frac{-x^2}{\tilde{\xi}^2}}+\Theta_0^2 \label{eqn:2Dto1DInitial}\ea in which we define $\tilde{t} = t - t_w$ as the time reference from the moment, $t_w$, when $\Theta_e$ is becomes uniform along the $x=0$ center.

Applying this new initial condition, we arrive at the quasi-1D solution:
\ba \tilde{\Theta}_e(x,\tilde{t}) = \left[\Theta_0^2 + e^{-\frac{\tilde{t}}{\tau_{\text{e-ph}}}} e^{\frac{-x^2}{\tilde{\xi}^2+4\mathcal{D}\tilde{t}}}\frac{\tilde{\Theta}_{hot}^2}{\sqrt{1+4\mathcal{D}\tilde{t}/\tilde{\xi}^2}}\right]^{1/2}.\label{eqn:2Dto1DSolution}\ea Analogous to the 1D solution, $\tilde{\xi}$ plays the role of the finite graphene flake width. Fig.~\ref{fig:2Dto1D}b compares the 1D and quasi-1D solutions for $w = 1~\mu$m and infinite $\tau_{\text{e-ph}}$. Specifically, we are interested in $\tilde{\Theta}_e$ for $t \gg \tilde{\xi}^2/4\mathcal{D}$ when the heat has diffused far enough from the hotspot ($t>t_w$ in \ref{fig:2Dto1D}a). In this case, we can compute the ratio of the quasi-1D and 1D temperature and find:
\ba \frac{\tilde{\Theta}_e}{\Theta_e} \simeq e^{\frac{t_w}{2\tau_{\text{e-ph}}}}\frac{\tilde{\Theta}_{hot}}{\Theta_{\text{hot}}}\sqrt{\frac{\tilde{\xi}}{\xi}} =\sqrt{\frac{\xi}{\tilde{\xi}}}\label{eqn:CalebsGoldenRatio}\ea Therefore, $\Theta_e$ and $\tilde{\Theta}_e$ maintain nearly a constant ratio as shown in Fig. \ref{fig:2Dto1D}b. The quasi-1D approximate solution in Eqn.\ \ref{eqn:2Dto1DSolution} suggests that, for $t \gg 4\mathcal{D}/\tilde{\xi}^2$, the finite width $w$ will effectively degrade $\Theta_e(x, t)$ by a geometric factor of $\sqrt{\xi/\tilde{\xi}}$ if the dissipation within the time period $t_w$ is negligible.

The $\tilde{\Theta}_e/\Theta_e$ ratio is also helpful to estimate the effect of finite graphene flake width on the detector efficiency as a function of the photon energy that we calculated and plotted in Fig. \ref{fig:efficiency}. Although the time integral to calculate the Josephson junction switching probability (Eqn. \ref{eqn:switching probability}) depends on details of the rise and fall of $\tilde{\Theta}_e$, we can approximate its degradation due to finite width by observing that for time $t\gg\tilde{\xi}^2/\mathcal{D}$, \ba \tilde{\Theta}_e \simeq \Theta_e\sqrt{\frac{\xi}{\tilde{\xi}}} \propto \Theta_{\text{hot}}\sqrt{\frac{\xi}{\tilde{\xi}}} \propto hf\sqrt{\frac{\xi}{\tilde{\xi}}} \label{eqn:lambdashiftapprox}\ea
based on Eqn. \ref{eqn:hotspot2} and \ref{eqn:uniform T} with $\Theta_{\text{hot}}\gg\Theta_0$. This suggests that we can interpret the finite width effect as a reduction of $\Theta_e$ (Fig.~\ref{fig:2Dto1D}c) and photon energy by the same geometric factor, $\sqrt{\xi/\tilde{\xi}}$. As a result, the detector efficiency for a finite width would simply shift the calculated results in Fig. \ref{fig:efficiency}c to the left-hand direction by $\sqrt{\xi/\tilde{\xi}}$. In reality, it is an experimental task to fabricate a narrow enough graphene absorber while maintain a high $\tau_{\text{e-ph}}$ in order to maximize the detector efficiency.

\begin{figure}[b]\centering
\includegraphics[width=0.9\columnwidth]{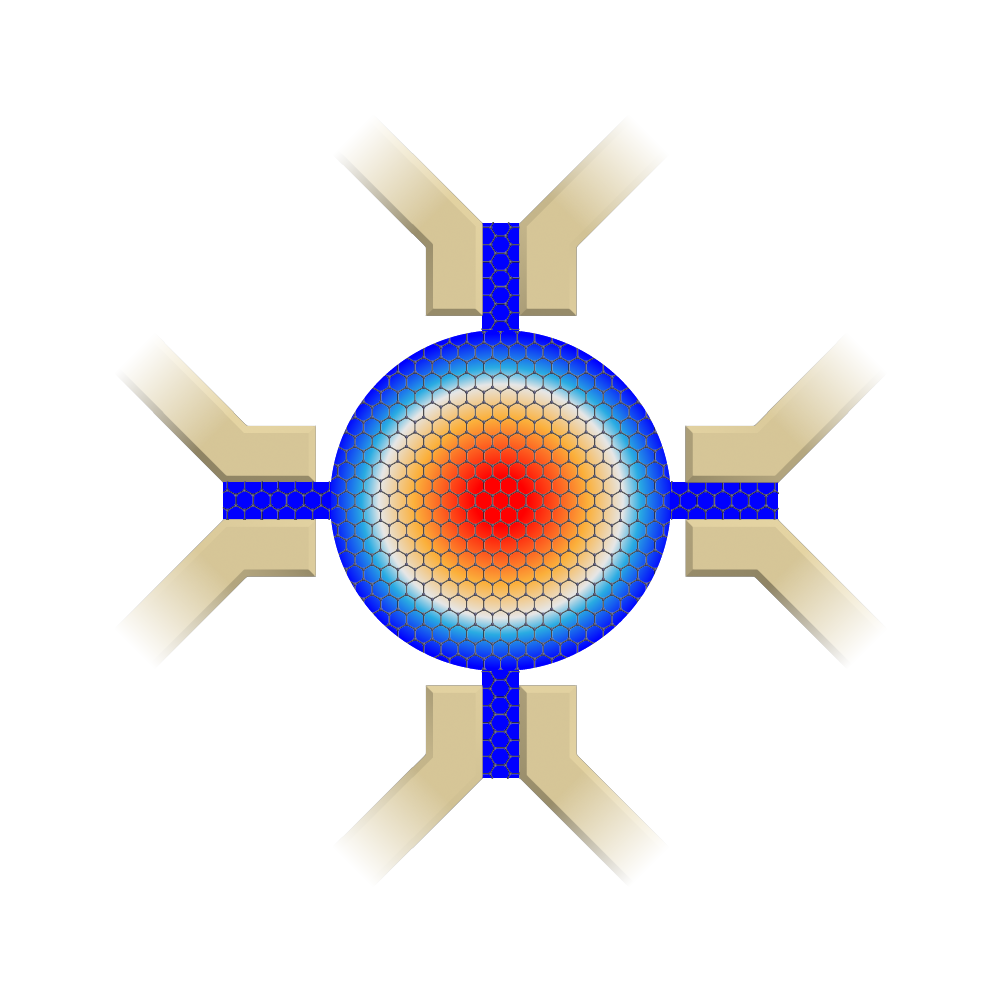}
\caption{\textbf{Device concept of a spatial-mode-resolving detector.} Our result points towards the idea that we can exploit the difference in the rise and fall of the electron temperatures at different locations on the graphene flake to resolve the spatial mode of a single photon. Time-resolved measurements of the Josephson junction sensors along the perimeter of the graphene absorber may resolve the location of photon absorption and thus the spatial mode.}
\label{fig:SpatialMode}
\end{figure}

\section{Outlook and conclusion}

The analysis of the dissipative heat diffusion model affirms the feasibility of simultaneously having efficient photon coupling, low-loss heat diffusion, and high-sensitivity superconductor-based readout in an IR graphene bolometer. While the temperature rise and fall in graphene as a function of position and time depends entirely on the detailed parameters as shown in Fig. \ref{fig:delta34size}, we find that a few simple formulas in solving the linearized differential equation can provide a basic understanding of thermal diffusion in the presence of dissipation. Detector performance also depends on the rate of heat dissipation, which varies between e-ph coupling regimes, i.e. supercollision, clean, or resonant scattering. Overall, our device modeling favors the supercollision regime when a longer integration time is needed, for instance to latch a Josephson junction to the normal state. Otherwise, the clean or resonant scattering e-ph coupling regimes may provide faster reset times due to their faster thermal relaxation in the non-linear regime. To use our numerical results for a different graphene strip width, we can use a simple geometric factor to find out the new detector efficiency.

A better understanding of thermal diffusion and dissipation can lead to photon detectors with expanded functionality. For instance, in Fig. \ref{fig:SpatialMode} we show a device concept for a spatial-mode-resolving (SMR) detector. With the photon coupling at the center of the graphene flake, comparison of the signals detected by multiple superconducting readouts along the circumference might be able to resolve the spatial mode of the incident photon. This concept is analogous to temporal sensing in SNSPDs, where the location of the hotspot in the nanowire can be detected by two amplifiers at opposite ends. Compared to the recent demonstration of SNSPD camera\cite{Oripov.2023}, graphene-based SMR detectors may have a larger spectral bandwidth for longer-wavelength photons. As a single photon can carry a large amount of quantum information by exploiting the multiple degrees of freedom in its spatial, temporal, and polarization modes, an SMR detector could provide a much needed technical capability in computation and communication based on single-photon quantum logic.

\section{Acknowledgement}
We thank J.~Balgley, M.~Kreidel, and J.~Park for the discussions. BJR and EAH acknowledge support under NSF CAREER DMR-1945278 and the Center for Quantum Leaps, Washington University in St.\ Louis. BH acknowledges that this research was supported by an appointment to the Intelligence Community Postdoctoral Research Fellowship Program at the Massachusetts Institute of Technology, administered by Oak Ridge Institute for Science and Education through an interagency agreement between the U.S. Department of Energy and the Office of the Director of National Intelligence. EGA acknowledges support from the Army Research Office MURI (Ab-Initio Solid-State Quantum Materials) Grant no. W911NF-18-1-043. GHL was supported by ITRC program (IITP-2022-RS-2022-00164799) and NRF grant (Nos. 2021R1A6A1A10042944, 2022M3H4A1A04074153 and RS-2023-00207732) funded by the Ministry of Science and ICT. This work is funded in part by NRO contract NRO000-20-C-0256. All findings and opinions are those of the authors and not of the U.S. government.
\bibliographystyle{naturemag}
\bibliography{pnrdfinal}
\end{document}